\documentclass[a4paper,fleqn,usenatbib]{mnras}

\usepackage{newtxtext,newtxmath}
\usepackage[T1]{fontenc}
\usepackage{ae,aecompl}

\usepackage{graphicx}	% Including figure files
\usepackage{amsmath}	% Advanced maths commands
\usepackage{amssymb}	% Extra maths symbols

\title[Ammonia in pre-stellar cores]{Why does ammonia not freeze out in the center of pre-stellar cores?}

\author[O. Sipil\"a, P. Caselli, E. Redaelli,  M. Juvela \& L. Bizzocchi]{
O. Sipil\"a$^{1}$\thanks{E-mail: osipila@mpe.mpg.de},
P. Caselli$^{1}$,
E. Redaelli$^{1}$,
M. Juvela$^{2}$
and L. Bizzocchi$^{1}$
\\
$^{1}$Max-Planck-Institute for Extraterrestrial Physics (MPE), Giessenbachstr. 1, 85748 Garching, Germany\\
$^{2}$Department of Physics, P.O.Box 64, FI-00014, University of Helsinki, Finland\\
}

\date{Accepted XXX. Received YYY; in original form ZZZ}

\pubyear{2018}

\begin{document}
\label{firstpage}
\pagerange{\pageref{firstpage}--\pageref{lastpage}}
\maketitle

\begin{abstract}
We carried out a parameter-space exploration of the ammonia abundance in the pre-stellar core L1544, where it has been observed to increase toward the center of the core with no signs of freeze-out onto grain surfaces. We considered static and dynamical physical models coupled with elaborate chemical and radiative transfer calculations, and explored the effects of varying model parameters on the (ortho+para) ammonia abundance profile. None of our models are able to reproduce the inward-increasing tendency in the observed profile; ammonia depletion always occurs in the center of the core. In particular, our study shows that including the chemical desorption process, where exothermic association reactions on the grain surface can result in the immediate desorption of the product molecule, leads to ammonia abundances that are over an order of magnitude above the observed level in the innermost 15000~au of the core -- at least when one employs a constant efficiency for the chemical desorption process irrespective of the ice composition. Our results seemingly constrain the chemical desorption efficiency of ammonia on water ice to below 1\%. It is increasingly evident that time-dependent effects must be considered so that the results of chemical models can be reconciled with observations.
\end{abstract}


%%%%%%%%%%%%%%%%%%%%%%%%%%%%%%%%%%%%%%%%%%%%%%%%%%

%%%%%%%%%%%%%%%%% BODY OF PAPER %%%%%%%%%%%%%%%%%%

\section{Introduction}

Ammonia is observed almost ubiquitously in the interstellar medium (ISM). It serves as a useful tool for measuring the kinetic gas temperature because of its particular spectroscopic properties, and hence understanding its chemical evolution allows us to deduce important information on physical processes in the ISM. The gas-phase chemistry of ammonia is well understood, but its evolution on the surfaces of interstellar dust grains is rather poorly constrained. Furthermore, the strength of ammonia desorption from the grain surfaces, and the nature of the desorption mechanism, are still open questions.

Chemical models of star-forming regions predict that, at low temperature ($T \sim 10 \, \rm K$), ammonia freezes out onto grain surfaces already at medium densities of a few times $10^5 \, \rm cm^{-3}$, which is attributed to its high binding energy \citep{Aikawa12,Taquet14,Sipila15b,Hily-Blant18}. These results are not in agreement with observations. Ammonia depletion has been observed toward pre-stellar cores, but it seems to occur only at very high (column) densities \citep{Friesen09, Ruoskanen11,Chitsazzadeh14}. To add to the conundrum, \citet{Crapsi07} derived an ammonia abundance profile in L1544, a well-studied pre-stellar core in Taurus, that increases toward the dust peak and shows no signs of depletion in the center of the core despite the high gas density ($n(\rm H_2) > 10^6 \, \rm cm^{-3}$).

The discrepancy between the modelling results and observations is puzzling given that ammonia is a relatively simple molecule and its main formation and destruction pathways consist of a small number of reactions. This problem must be investigated in detail so that our understanding of the gas-grain chemical interaction can be improved. To this end, we used a comprehensive gas-grain chemical model to simulate the abundance of gas-phase ammonia in L1544. We varied several modelling parameters that are a priori expected to influence the ammonia abundance, in an effort to produce solutions where ammonia depletion either does not occur, or is mitigated to the observed levels within the uncertainties.

The paper is organized as follows. Sect.\,\ref{s:model} discusses our chemical code and recent updates to it. Here we also discuss the physical source models used in the paper, and our parameter-space approach to the modelling. In Sect.\,\ref{s:results} we present our results which are discussed in Sect.\,\ref{s:discussion}. We give our conclusions in Sect.\,\ref{s:conclusions}. Additionally, a benchmark of radiative codes is presented in Appendix~\ref{a:benchmark}.

\section{Model}\label{s:model}

\subsection{Chemical model}\label{ss:chemicalcode}

We used an expanded version of the gas-grain chemical code described in \citet{Sipila13,Sipila15a,Sipila15b}. In short, the code solves a system of rate equations connecting separate networks of gas-phase and grain-surface chemistry. The details of the fundamental chemical processes (e.g., adsorption, thermal and non-thermal desorption) including the relevant formulae are described in \citet{Sipila15a} and are not reproduced here for the sake of brevity. However, for the purposes of the present paper, it was necessary to expand the set of chemical processes considered in the code, as opposed to using the older model as described in \citet{Sipila15a}. We list these additions below.

{\sl Cosmic-ray induced secondary photoreactions.} The interiors of molecular clouds are well shielded from the ultraviolet (UV) photons prominent in the spectrum of the interstellar radiation field (ISRF). However, cosmic-ray(CR)-induced ionization of $\rm H_2$ followed by electron recombination can create an UV field of appreciable strength inside otherwise well-shielded regions \citep{Sternberg87,Gredel89}. This UV field may ionize and/or dissociate molecules in the gas phase and on the surfaces of grains. The rate coefficient for the ionization or dissociation of atom or molecule $i$ in this process is given by
\begin{equation}\label{eq:secphot}
k_{\rm secphot}(i) = \zeta_p({\rm H}) X({\rm H_2})\frac{p_i}{1-\omega} ~ [\rm s^{-1}] \, ,
\end{equation}
where $\zeta_p({\rm H})$ is the primary CR ionization rate per hydrogen atom, $X({\rm H_2}) = n({\rm H_2})/n_{\rm H}$ is the fractional abundance of ${\rm H_2}$ ($n_{\rm H}$ is the total number density of hydrogen nuclei), $p_i$ is an efficiency factor for the ionization/dissociation reaction in question, and $\omega$ is the grain albedo (assumed $\omega = 0.5$). We updated the list of reactions and efficiency factors included in the current release of the KIDA network (\citealt{Wakelam15}; see below) using the data of \citeauthor{Heays17}\,(\citeyear{Heays17}; their Table~20). The efficiency factors required here were obtained by simply dividing their photoionization/dissociation rates by $10^{-16}$ (see also \citealt{Hily-Blant18}). As noted by \citet{Heays17}, the simple division of the rates leads only to an approximate agreement with the formalism of \citet{Gredel89}, but a more detailed treatment of this issue is beyond the scope of the present paper. We also note that the factor $X({\rm H_2})$ in Eq.\,(\ref{eq:secphot}) was originally missing in the work of \citet{Gredel89}, but is necessary for the present context as pointed out by \citet{Woodall07} and \citet{Flower07}.

{\sl Chemical desorption.} Two-body chemical reactions with one reaction product on the grain surface may lead to the immediate desorption of the reaction product, if the excess formation energy is absorbed by the grain \citep{Williams68, Watson72a, Watson72b}. This process is commonly referred to as chemical desorption, or reactive desorption. Different treatments of chemical desorption in the context of gas-grain models have been suggested in the literature. Here we adopt the approach of \citet{Garrod07}, in which exothermic surface reactions lead to desorption with a probability of $\sim$1\%. Recent investigations \citep{Dulieu13, Minissale16a, Chuang18} have shown that the efficiency of chemical desorption may vary significantly depending on the reaction and type of surface. These results have already been incorporated in chemical models \citep{Vasyunin17}. However, because of the high uncertainties involved, we chose to follow the uniform $\sim$1\% approach of \citet{Garrod07}.

{\sl $\rm H_2$ self-shielding.} In molecular clouds, and in particular in starless and pre-stellar cores, hydrogen exists mostly in molecular form, and the $\rm H_2$ present in the outer cloud may efficiently shield the $\rm H_2$ in the inner cloud against UV radiation. This effect is important because of the role that H atoms play in surface chemistry. Also, $\rm H_2$ self-shielding affects the $\rm H_2$ ortho/para ratio \citep{Sipila18}, which will in turn affect the chemistry of ammonia because the $\rm N^+ + H_2 \longrightarrow NH^+ + H$ reaction that initiates ammonia formation is strongly endothermic in the presence of para-$\rm H_2$ \citep{Dislaire12}. We adopted the $\rm H_2$ self-shielding factor from \citet{Draine96}. This factor depends on the total $\rm H_2$ column density (see Sect.\,\ref{ss:physmodel} for details on the physical core model used here) and is included in the rate equations that involve the ionization or dissociation of (ortho or para) $\rm H_2$. In reality self-shielding applies to other abundant molecules such as CO \citep{Visser09} and $\rm N_2$ \citep{Heays14} as well, but the self-shielding factor is a function of column density which for these species is a highly time-dependent quantity\footnote{Unlike that of $\rm H_2$, which we assume to remain constant as we consider a static physical model; see Sect.\,\ref{ss:physmodel}.}. Self-shielding is thus not included for species other than $\rm H_2$ because our model is not fully time-dependent (see Sect.\,\ref{ss:physmodel}). The limited inclusion of self-shielding does not affect the results presented in this paper as we concentrate on the inner, heavily shielded, areas of L1544.

{\sl Temperature-dependent sticking coefficients.} One of the parameters controlling the adsorption of molecules onto dust grains is the sticking coefficient. Many chemical models assume that the sticking coefficients of the various species equal unity regardless of the temperature of the medium, and indeed this assumption was made in our previous models as well. For the present work, we updated the chemical model to include temperature-dependent sticking coefficients for selected species, namely H, $\rm H_2$, $\rm N_2$, CO, $\rm O_2$, $\rm CH_4$, and $\rm CO_2$, including deuterated variants whenever applicable. For atomic H, we use the parametrized expression from \citet{Cuppen10}. Their formula applies to a graphite surface, but the sticking coefficient on water ice is almost identical to that on graphite for temperatures below 100\,K \citep{Cuppen10}. We assume the same formula for atomic D. For the rest of the species listed above we adopted the sticking coefficients presented by \citet{He16}, who also derived a sticking coefficient for $\rm D_2$. For $\rm HD$ and deuterated methane we assume that the sticking coefficient equals that of $\rm H_2$ or $\rm CH_4$, respectively. Species for which there is no theoretical or experimental data are assumed to stick with an efficiency of unity.

The additions to the chemical model presented above are essential because they either directly or indirectly affect the abundance of ammonia, which is the main target of our simulations.

Our gas-phase reaction network is essentially the kida.uva.2014 network \citep{Wakelam15} that was deuterated and spin-state separated according to the prescriptions laid out in \citet{Sipila15a,Sipila15b}, and updated as described above. A similar procedure was applied to our base grain-surface network, which is an updated version of the one presented by \citet{Semenov10}. For reactions involving CRs or photons we assume the same rate coefficients in the gas phase and on the grain surface, with the important distinction that we only consider dissociation reactions on the grain surface (we assume that no ionic species exist in the ice). Photodesorption is also included for a limited set of species \citep{Sipila18}. The secondary photoreactions discussed above were added to both the gas-phase and grain-surface networks, again ignoring pathways that produce ions on the grain surface. The gas-phase and grain-surface networks contain a combined total of $\sim$82000 reactions.

\subsection{Physical model for L1544}\label{ss:physmodel}

\begin{figure}
	\includegraphics[width=\columnwidth]{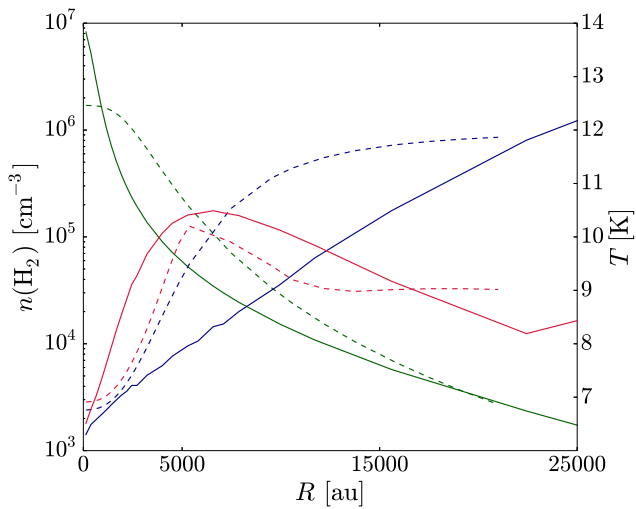}
    \caption{Density and temperature profiles for L1544 used in this work. Solid lines display the profiles from \citet{Keto10a}: density (green), dust temperature (blue), and gas temperature (red). The dashed lines show the corresponding data from an alternative model (see text) based on \citet{Chacon-Tanarro19}.}
    \label{fig:physicalmodels}
\end{figure}

By default, we use the density and (gas and dust) temperature structure from the one-dimensional (1D) L1544 source model published by \citeauthor{Keto10a}\,(\citeyear{Keto10a}; hereafter K10; see also \citealt{Keto14}). This model has been previously used for interpreting observations and for carrying out chemical modelling of L1544 in several studies, such as \citet{Bizzocchi13}, \citet{Sipila16a}, \citet{Vasyunin17}. The density and temperature structures of the model core are plotted in Fig.\,\ref{fig:physicalmodels}.

However, recent observations by \citeauthor{Chacon-Tanarro19}~(\citeyear{Chacon-Tanarro19}; hereafter C19) show that the central density of L1544 may in fact be a factor of $\sim$4 lower than that calculated by K10, and that the slope of the density profile outside the flat radius may be different. In C19 a new dust temperature profile for L1544 was also presented. These new density and dust temperature profiles are plotted in Fig.\,\ref{fig:physicalmodels} along with the profiles of K10, and are used in one of the models introduced below (see Sect.\,\ref{ss:parameterspace}). Recent dust continuum emission observations toward L1544 obtained with the Atacama Large (sub)Millimeter Array agree with the profiles derived by C19, but local density enhancements up to $\sim10^7 \, \rm cm^{-3}$ are also present within the central 1400\,au \citep{Caselli19}.

For the purposes of the present work, a gas temperature profile is required as well. To this end, we took the density and dust temperatures from C19 and used our hydrodynamical code \citep{Sipila18}, with infall/expansion velocity forced to zero, to calculate a (representative) gas temperature, plotted in Fig.\,\ref{fig:physicalmodels} as the red dashed line. The profile corresponds to $t = 10^5\,\rm yr$ of chemical evolution, which is a reasonable timescale since the physical model is static and the abundances of the cooling molecules hardly change after $t = 10^5\,\rm yr$ in the inner part of the core, which is our main interest presently, owing to the high density. We only used this alternative L1544 physical structure in one of our model runs, the details of which are explained in Sect.\,\ref{ss:parameterspace}.

\begin{table}
	\centering
	\caption{Initial abundances (with respect to the total hydrogen number density $n_{\rm H}$) used in the chemical modelling. The initial $\rm H_2$ ortho/para ratio is $1.0 \times 10^{-3}$.}
	\begin{tabular}{l|l}
		\hline
		Species & Abundance\\
		\hline
		$\rm H_2$ & $5.00\times10^{-1}$\\
		$\rm He$ & $9.00\times10^{-2}$\\
		$\rm HD$ & $1.60\times10^{-5}$\\
		$\rm C^+$ & $1.20\times10^{-4}$\\
		$\rm N$ & $7.60\times10^{-5}$\\
		$\rm O$ & $2.56\times10^{-4}$\\
		$\rm S^+$ & $8.00\times10^{-8}$\\
		$\rm Si^+$ & $8.00\times10^{-9}$\\
		$\rm Na^+$ & $2.00\times10^{-9}$\\
		$\rm Mg^+$ & $7.00\times10^{-9}$\\
		$\rm Fe^+$ & $3.00\times10^{-9}$\\
		$\rm P^+$ & $2.00\times10^{-10}$\\
		$\rm Cl^+$ & $1.00\times10^{-9}$\\
		\hline
	\end{tabular}
	\label{tab:initialabundances}
	%\tablefoot{Footer}
\end{table}

As in our previous works \citep[e.g.,][]{Sipila16a}, we derived radius-and-time-dependent chemical abundance profiles in L1544 by dividing the physical model into concentric shells, calculating the chemical evolution separately in each shell, and combining the results obtained in the different shells at a given time step. We assume that the gas is initially atomic with the exception of $\rm H_2$ and HD; the adopted initial abundances are presented in Table~\ref{tab:initialabundances}. These initial abundances are used for all of the models presented in this paper except when otherwise noted (see below).

\subsection{Parameter-space exploration}\label{ss:parameterspace}

\begin{table*}
	\centering
	\caption{Parameter variations and the associated model denominations considered in this paper. Our standard model is highlighted in boldface.}
	\begin{tabular}{lccr}
		\hline
		Model grid (total of 108 models) \\
		\hline
		Parameter & & Values considered \\
		\hline
		Binding energy of $\rm NH_3$ (K) & {\bf 5500} & 3000 & 1000 \\
		Sticking coefficient of atomic N & {\bf 1} & 0.3 \\
		Photodesorption (of CO, $\rm (o/p) H_2O$, $\rm CO_2$, $\rm N_2$) & {\bf Yes} & No \\
		Photodesorption of $\rm (o/p) NH_3$ & Yes & {\bf No} \\
		Chemical desorption & {\bf Yes} & No \\
		External $A_{\rm V}$ (mag) & 5 & 2 & {\bf 1} \\
		\hline
		Single models (see text for further explanations) \\
		\hline
		S1: Low elemental N abundance \\
		S2: Multilayer ice chemistry \\
		S3: Initial abundances from a lower-density cloud model \\
		S4: C19 density profile \\
		%S5: Modified rate coefficients for surface chemistry; H and D diffusion by tunneling \\
		S5: Gas-phase chemistry only \\
		S6: Initial $\rm H_2$ o/p ratio of 0.1 \\
		S7: CR ionization rate of $\zeta_p = 10^{-16} \, \rm s^{-1}$ \\
		S8: As S7, but no chemical desorption \\
		S9: Grain radius $a_{\rm g} = 0.2 \, \mu \rm m$ \\
		S10: Grain radius $a_{\rm g} = 0.05 \, \mu \rm m$ \\
		S11: Hydrodynamical model \\
		\hline
	\end{tabular}
	\label{tab:models}
\end{table*}

The chemistry of ammonia is sensitive to various model parameters. For example, disregarding chemical desorption will have a direct impact on the ammonia abundance; the binding energy of ammonia is very high, and without efficient non-thermal mechanisms it will, effectively, not desorb at low temperature (close to 10\,K). To investigate the sensitivity of the ammonia abundance on the model assumptions, we constructed a grid of models in which we varied several key parameters that influence the ammonia abundance, particularly on the grain surfaces. The parameters and their variations are displayed in Table~\ref{tab:models}. Our fiducial model is highlighted in boldface.

The binding energy of ammonia on water ice is uncertain. Our fiducial value of 5500\,K is taken from \citet{Collings04}. However, lower values of the order of 3000\,K have been suggested \citep[see][and references therein]{Kamp17}. We consider both values in our model grid. Furthermore, as a test, we also consider a very low (ad hoc) value of 1000\,K to explore its effect on the ammonia abundance.

We take a fiducial value of unity for the sticking coefficient of atomic N, as is most often assumed in astrochemical models. The possibility of lower values has been suggested by \citet{Flower06b}, and we take a value of 0.3 as an alternative to the canonical value of unity.

By default, we consider the photodesorption of CO, (o/p)$\rm H_2O$, $\rm CO_2$, and $\rm N_2$. The assumed photodesorption yields are, respectively, $2.7 \times 10^{-3}$ \citep{Oberg09a}, $1.0 \times 10^{-3}$ \citep{Oberg09b}, $1.2 \times 10^{-3}$ \citep{Oberg09a}, and $3.0 \times 10^{-4}$ \citep{Oberg09a}. The $\rm CO_2$ yield only approximates the complex expression derived by \citet{Oberg09a} that depends on the ice thickness, which we do not track in this paper. We assume that both ortho and para $\rm H_2O$ are photodesorbed with the same yield. We also tested cases where $\rm NH_3$ photodesorption is included, with a yield of $1.0 \times 10^{-3}$ \citep{Martin-Domenech18}. This yield is assumed to apply to both ortho and para $\rm NH_3$, and we note that the experimental yield has been derived for pure $\rm NH_3$ ice, and not for $\rm NH_3$ on water or an ice mixture.

We assume that the core model is embedded in a larger molecular cloud; the parameter ``external $A_{\rm V}$'' expresses the attenuation (in the visual) by the parent cloud. All together, the grid consists of 108 parameter combinations.

We also ran eleven single models separately from the parameter grid in order to test some additional assumptions. These models, also collected in Table~\ref{tab:models}, were run as separate cases mainly because of calculational time constraints. It takes roughly three hours to complete one model run on a standard desktop computer (with parallelization), and every new parameter change in the grid would double the required computational time. The use of supercomputing resources for the present work is however not necessary as we are not searching for an exact fit to available observational data, and the main interest is in investigating the general trends caused by the parameter variations.

The single models S1 to S10 correspond to our fiducial model with the particular changes indicated in Table~\ref{tab:models}. In model~S1, we assumed an N elemental abundance of $2.14\times10^{-5}$ (\citealt{Wakelam08}; their model EA1). Model~S2 incorporates a multilayer (three-phase) ice model as detailed in \citet{Sipila16b}\footnote{Although this model naturally tracks the ice thickness, we still used the approximate $\rm CO_2$ photodesorption yield noted above.}. For model~S3, we first calculated a single-point chemical model with $n(\rm H_2) = 5 \times 10^3 \, \rm cm^{-3}$, $T_{\rm gas} = T_{\rm dust} = 15\,\rm K$, $A_{\rm V} = 1 \, \rm mag$, and extracted the abundances from that model at $t = 10^5 \, \rm yr$ to use as initial abundances for the fiducial model. Model~S4 uses the C19 density profile for L1544 (see Sect.\,\ref{ss:physmodel}) instead of the Keto et al. profile. Model~S5 considers gas-phase chemistry only, where the formation of $\rm H_2$, HD, and $\rm D_2$ is parametrized as in \citet{Kong15}. In model~S6 we adopt an initial $\rm H_2$ ortho/para ratio of 0.1. Models S7 and S8 incorporate an increase of the CR ionization rate over our standard value of $\zeta_p = 1.3 \times 10^{-17} \, \rm s^{-1}$, and the latter model also excludes chemical desorption. In models S9 and S10 we multiply or divide our standard grain radius (0.1\,$\mu$m) by a factor of two.

Finally, model~S11 represents a hydrodynamical model calculation, the details of which are given in Sect.\,\ref{ss:hydro}.

\section{Results}\label{s:results}

We extracted the abundance profiles of ortho and para $\rm NH_3$ at different time steps from the models detailed in Table~\ref{tab:models}, and compared the results against the observations of \citet{Crapsi07}. We present the results of the comparison in what follows, broken down into the results from our parameter grid and the single models S1 to S11.

\subsection{Parameter grid}

\begin{figure}
	\includegraphics[width=\columnwidth]{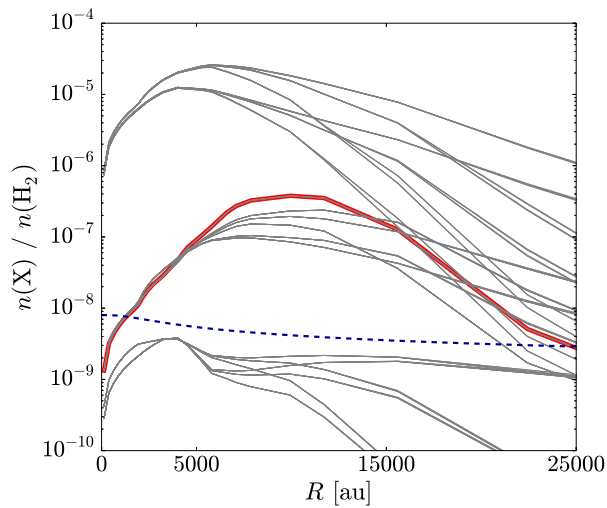}
    \caption{Abundance of ortho+para $\rm NH_3$ as a function of radius, convolved to a beam of 4\arcsec. The results from our parameter grid (108 models in total) are displayed as gray solid lines. The red solid line represents our fiducial model, and the dark blue dashed line represents the abundance profile deduced by \citet{Crapsi07}. The model results correspond to $t = 10^5\,\rm yr$.}
    \label{fig:paramgrid}
\end{figure}

Figure~\ref{fig:paramgrid} displays the abundance profiles of ortho+para $\rm NH_3$ obtained from all of the models comprising our parameter grid, convolved to a beam size of 4$\arcsec$ for comparison with the observations by \citet{Crapsi07}. The abundance profile from \citet{Crapsi07} is also shown for comparison. One property of the calculated models is immediately evident: we always obtain solutions where the ammonia abundance depletes at the center of the core, and the monotonically inward-increasing profile from Crapsi et al. is never reproduced.

\begin{figure*}
	\includegraphics[width=2.0\columnwidth]{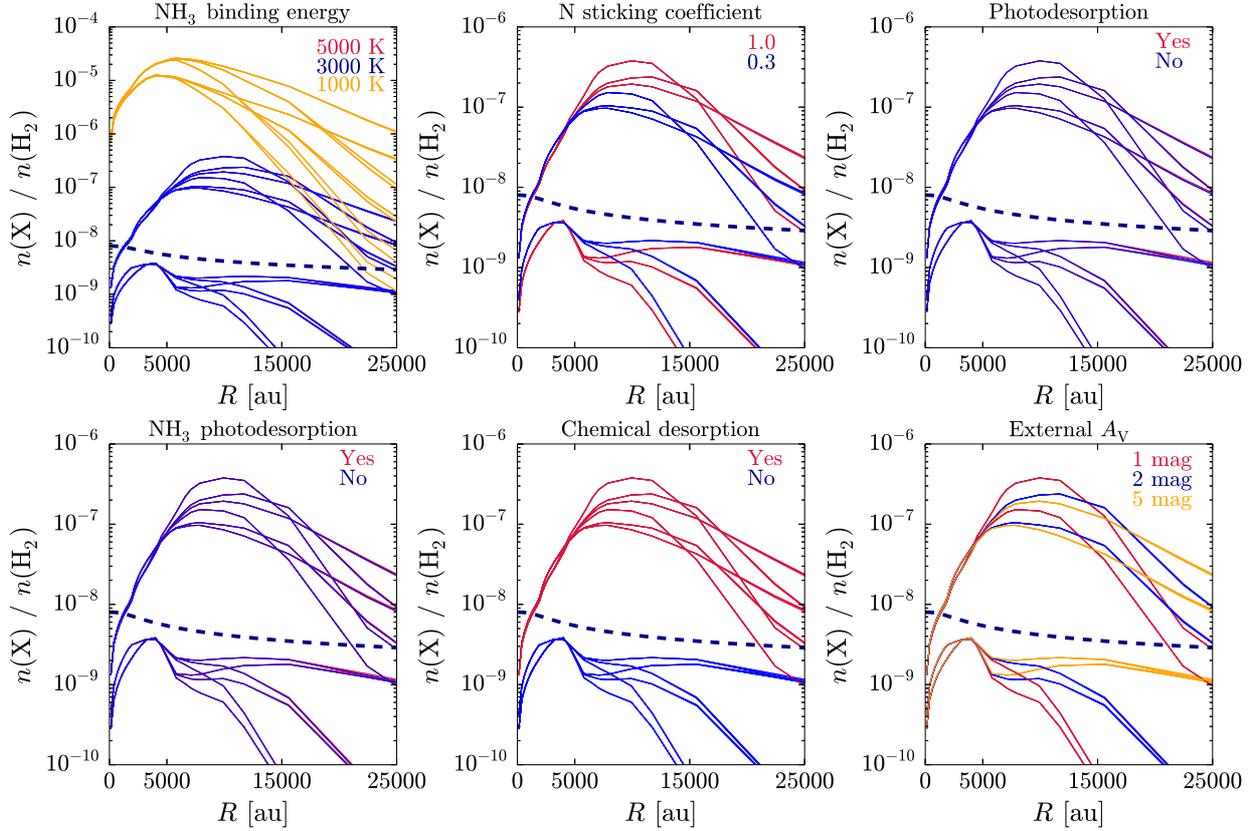}
    \caption{Data presented in Fig.\,\ref{fig:paramgrid} but with different parameter cases separated and highlighted. {\sl Upper left:} $\rm NH_3$ binding energy 5000\,K (red), 3000\,K (blue; overlaps with red), or 1000\,K (orange). {\sl Upper middle:} atomic N sticking coefficient of unity (red) or 0.3 (blue). {\sl Upper right:} photodesorption (without $\rm NH_3$) on (red) or off (blue). {\sl Lower left:} $\rm NH_3$ photodesorption on (red) or off (blue). {\sl Lower middle:} chemical desorption on (red) or off (blue). {\sl Lower right:} external $A_{\rm V}$ of 1\,mag (red), 2\,mag (blue), or 5\,mag (orange). The dark blue dashed lines represent the abundance profile deduced by \citet{Crapsi07}.}
    \label{fig:paramseparation}
\end{figure*}

Many of the model solutions are virtually identical to each other and overlap. Also, the solutions appear in distinct groups. It is therefore sensible to separate the solutions based on the individual parameter values. In Fig.\,\ref{fig:paramseparation}, we highlight the effect of each parameter. The effect of the $\rm NH_3$ binding energy is shown in the top left panel. It is immediately evident that the very low (ad hoc) value 1000\,K leads only to solutions where the modeled abundance is orders of magnitude above the observed one in the central areas of the core. These solutions can therefore be discarded, and are not displayed in any of the figures from here on. Notably, using a value of 3000\,K or 5000\,K has very little influence to the results (the red and blue lines overlap). The two remaining sets of curves correspond to models where chemical desorption is turned on or off, respectively (see the lower middle panel of Fig.\,\ref{fig:paramseparation}), and these sets are further subdivided as described below.

The sticking coefficient of atomic nitrogen does not influence the solutions in a clearly predictable manner, and solutions with both high and low overall $\rm NH_3$ abundances can be obtained with both of the tested values of the coefficient. Photodesorption, with or without $\rm NH_3$, has a negligible influence on our results. We note that \citet{Furuya18} have recently demonstrated that  ammonia photodesorption can have an impact on the gas-phase N and $\rm N_2$ abundances at low visual extinctions. Our model does not reproduce this effect. We performed test calculations which indicate that the effect of photodesorption is much greater in three-phase (i.e., the ice is separated into a mantle and a bulk) models than it is in the two-phase model that we consider here (except in model S2). Also, our description of ice chemistry is different from that of \citet{Furuya18}, which causes discrepancy in the modeling results (priv. comm. with K. Furuya). A detailed comparison is out of the scope of the present paper.

Chemical desorption influences the results greatly; we obtain solutions with very high peak ammonia abundances if chemical desorption is included, while the peak ammonia abundance never rises above a few~$\times~10^{-9}$ if chemical desorption is turned off. This is a very strong result given that we are using the conservative value of $\sim$1\% for the efficiency of the chemical desorption process \citep{Garrod07}, which is much lower than the values derived for some reactions by \citet{Minissale16a}, for example. Finally, the magnitude of external $A_{\rm V}$ changes the shape of the ammonia abundance profile; if external $A_{\rm V}$ is low, we obtain rather narrow abundance distributions, while increasing values of the external $A_{\rm V}$ yield increasingly extended distributions.

In conclusion to the above: our models do not reproduce the shape of the observed ammonia abundance profile, and we always obtain solutions where $\rm NH_3$ depletes near the core center. If we disregard the discrepancy in the central few thousand au, the best fit to the observations of \citet{Crapsi07} is reached with a model where {\it chemical desorption is excluded, the sticking coefficient of atomic N is unity, and the external $A_{\rm V}$ is higher than 2~mag}.

\subsection{Single models S1 to S10}\label{ss:singlemodels}

\begin{figure*}
	\includegraphics[width=1.8\columnwidth]{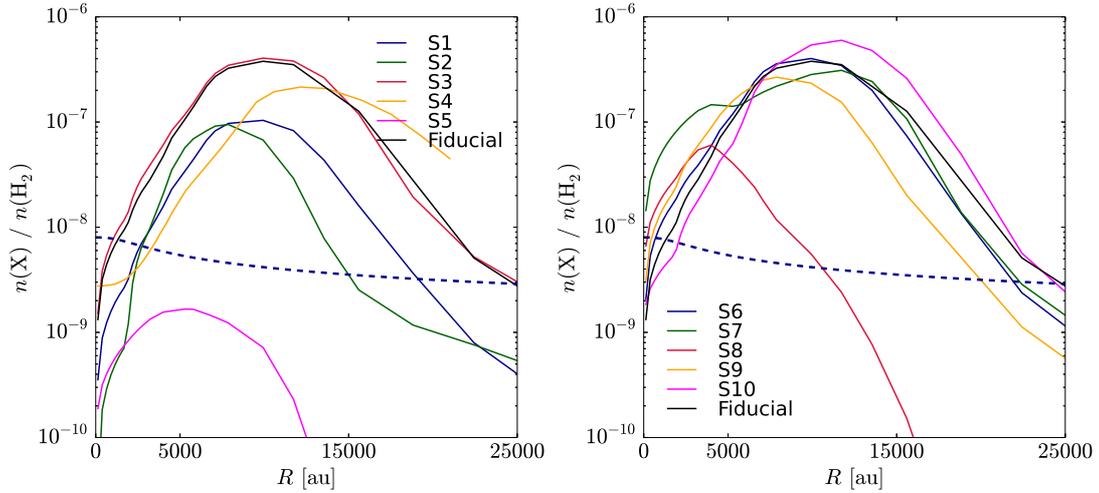}
    \caption{As Fig.\,\ref{fig:paramgrid}, but showing the single models S1 to S10 discussed in the text, and our fiducial model. The dark blue dashed line represents the abundance profile deduced by \citet{Crapsi07}.}
    \label{fig:singlemodels}
\end{figure*}

Figure~\ref{fig:singlemodels} shows the $\rm NH_3$ abundance profiles predicted by the single models S1 to S10. A significant spread is evident in the peak ammonia abundances depending on the model and, most notably, none of the single models provide a solution where ammonia depletion does not occur.

Model~S1 presents an ammonia abundance profile that is very similar to that given by our fiducial model, except scaled down. In model~S2, where a three-phase ice description is adopted, ammonia depletes very strongly because of trapping in the inert ice bulk beneath the active surface layer(s) \citep[for more details see][]{Sipila16b}. Furthermore, because of the trapping, the chemical desorption process is not efficient and we obtain a lower ammonia peak abundance than in the fiducial model. If the gas is first let to evolve in a diffuse cloud environment (model~S3), the end result is very close to the fiducial model, suggesting that {\it the initial conditions no longer play a role if the gas is let to evolve for a sufficient amount of time in the pre-stellar phase}.

When the density profile from C19 is used (model~S4), the ammonia depletion zone is larger than in our fiducial model. This is because the central high-density area, where ammonia depletes efficiently, is broader in the C19 model than in that of Keto et al., even though the density at the very center of the core is a factor of $\sim$4 lower in the former (see Fig.\,\ref{fig:physicalmodels}). The results from model~S4 and the fiducial model are again qualitatively similar despite the rather large difference in the density profiles.

The ammonia abundance profile does not display an inward-increasing trend even in the gas-phase model~S5. This effect is tied to the electron fraction. The production chain of ammonia starts with $\rm N^+$, which is produced in reactions between $\rm He^+$ and $\rm N_2$. At high density and low temperature, the rate of electron impacts with $\rm He^+$ is high, which inhibits the production of $\rm N^+$ (in contrast with gas-grain models where the abundance of $\rm He^+$ tends to increase with freeze-out). On the other hand $\rm N_2$ is not produced efficiently at low density, so that the formation of ammonia is (in a gas-phase model) the most efficient at a density of $\sim 10^5 \, \rm cm^{-3}$. We note that previous gas-phase models have predicted higher ammonia abundances of the order of $10^{-8}$ even at high density \citep[e.g.,][]{LeGal14,Roueff15}. These models however incorporated rate coefficients for some important reactions in the ammonia formation network that are much higher than the up-to-date values included in kida.uva.2014 and hence in the present model. The rate coefficient revisions lead to generally lower ammonia abundances, highlighting the great sensitivity of chemical models to uncertainties in input data.

In model~S6 we tested a higher initial $\rm H_2$ ortho/para ratio. Evidently, we obtain an abundance profile that is very similar to the fiducial model, providing further evidence of the insensitivity of the ammonia abundance to the initial conditions.

Models S7 and S8 explore the effect of the CR ionization rate. In addition to testing the effect of simply increasing the ionization rate (S7), we also tested a case where the chemical desorption process is additionally turned off (S8). Evidently, the enhancement in the CR ionization rate helps to maintain a higher abundance of ammonia in the central core, although depletion still occurs at the highest densities. However, the effect of CRs is so strong that it counteracts to some degree the exclusion of chemical desorption, and we obtain even in model S8 an ammonia abundance that is much higher than the observed one.

Finally, in models S9 and S10 we varied the grain radius. Increasing or decreasing the grain radius does not modify the abundance profile in a significant way as compared to the fiducial model, and the same trends as in the majority of our models are displayed here as well.

\subsection{Single model S11: hydrodynamics}\label{ss:hydro}

We have recently demonstrated that chemistry plays a very important role in determining the dynamics of the collapse of a star-forming cloud, and that using a static model for the physical structure of the core -- as in the present paper so far -- either overestimates or underestimates chemical abundances in a collapsing core, depending on both radius and time \citep{Sipila18}. That study was partly motivated by the previous observations of ammonia toward L1544, and therefore it is logical to consider if including dynamics could solve the present problem of excessively strong ammonia depletion appearing in the models, or at least to alleviate it.

To this end, we re-ran the hydrodynamical simulation described in \citet{Sipila18}, i.e., we started with an unstable Bonnor-Ebert sphere with central density $n_{\rm c}({\rm H_2}) = 2 \times 10^4 \, \rm cm^{-3}$, temperature $T = 10 \, \rm K$, and mass $M \sim 7.15 \, M_{\sun}$, but used the updated chemical networks described above. In \citet{Sipila18}, the termination condition of the hydrodynamical code was set to correspond to the time step when the infall flow becomes supersonic, and in that paper this occurred at $t = 7.19 \times 10^5 \, \rm yr$. In the present case the termination of the code occurred at $t = 1.20 \times 10^6 \, \rm yr$. There are three reasons for this difference. First, the changes to the chemical setup introduced here affect the chemistry at low density in particular, and this is reflected on the infall velocity profile and hence on the collapse timescale. Second, we have fixed a minor coding error in the expression that compares the infall velocity and sound speed to determine the termination time; the old version of the code calculated the sound speed inadvertently using the maximum of the gas temperature instead of its local value (the effect of this error is very small). Third, and most importantly, the cooling efficiency of HCN is clearly lower in the present paper than in \citeauthor{Sipila18}\,(\citeyear{Sipila18}; see also below). In fact, we have determined through extensive testing that the high HCN cooling efficiency presented in \citet{Sipila18} is a numerical error, the exact cause of which we have however not been able to identify. It remains unknown why only HCN was affected while the cooling powers of the other coolants are similar in both our current and earlier model calculations.

L1544 may be somewhat more massive than the Bonnor-Ebert sphere used here (for example K10 employed a Bonnor-Ebert sphere with mass $M~\sim~10 \, M_{\sun}$), but we chose to use the same initial core configuration as in \citet{Sipila18} in order to easily track down the causes for the differences in the physical and chemical evolution due to sources other than the initial physical model.

\begin{figure*}
	\includegraphics[width=2.0\columnwidth]{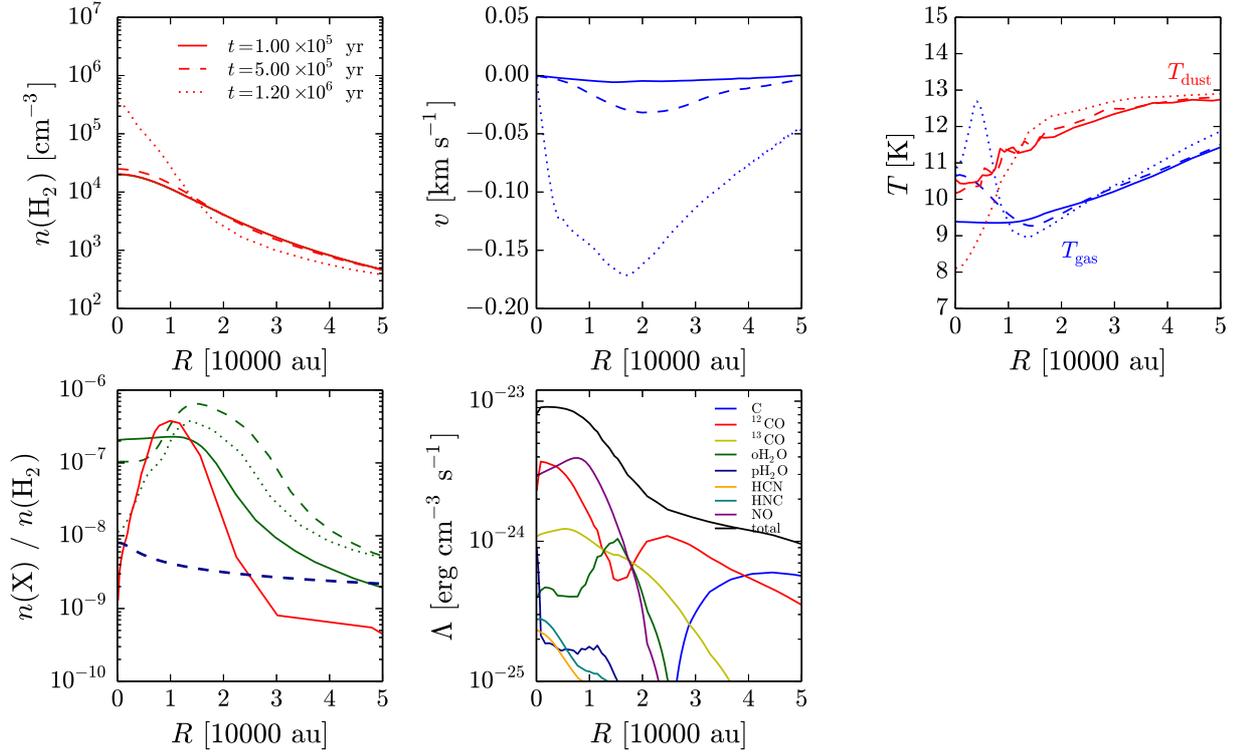}
    \caption{{\sl Upper left}: Density profile of the model core at three different time steps, indicated in the panel. {\sl Upper middle}: Infall velocity profile at the same three time steps. {\sl Upper right}: Gas (blue) and dust (red) temperature profiles at the same three time steps. {\sl Lower left}: Abundance profile of ortho+para ammonia at the same three time steps. Also shown are the result from our fiducial model at $t = 10^5 \, \rm yr$ (red), and the abundance profile deduced by \citet{Crapsi07} extrapolated to 50000\,au for illustration purposes (dark blue). {\sl Lower middle}: Cooling powers of selected molecules, indicated in the panel, at $t = 5 \times 10^5 \, \rm yr$ in the hydrodynamical simulation.}
    \label{fig:hydro}
\end{figure*}

The results of the hydrodynamical simulation are shown in Figure\,\ref{fig:hydro}. Unlike in the other abundance plots presented in this paper which concentrate on the inner 25000\,au, we plot here the results up to 50000\,au to facilitate easier comparison to Figs.\,3~and~4 of \citet{Sipila18}. First, it is strikingly evident that the HCN cooling efficiency is very low in the current model, while other cooling efficiencies are similar, as compared to \citet{Sipila18}. In particular, we still obtain a very strong contribution from NO in dense gas at late times. Comparison of the solution at the final time step ($t = 1.20 \times 10^6 \, \rm yr$) to the present fiducial (static) chemical model shows that ammonia is very strongly depleted even in the hydrodynamical model. The depletion factor is smaller than in the static model, but this is only because the central density is less than $10^6\,\rm cm^{-3}$ at the time of the termination of the calculation. If the calculation was continued beyond this point, ammonia would deplete as strongly as it does in the static case. The ammonia depletion zone is seemingly smaller in the static model, but this is only because the K10 physical model is more centrally concentrated (Fig.\,\ref{fig:physicalmodels}) than the hydrodynamical solution at the final time step. We note that the abrupt changes in some of the cooling functions near the origin are only transient radiative transfer artifacts that do not affect the overall evolution of the core.

\subsection{Time dependence}

\begin{figure*}
	\includegraphics[width=2.0\columnwidth]{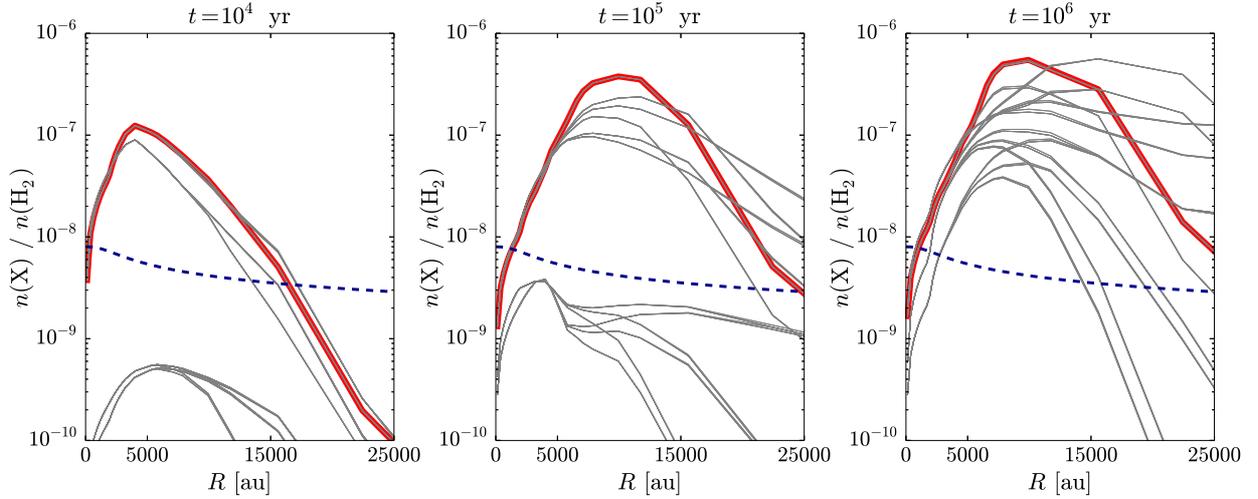}
    \caption{Ammonia abundance profiles from our model grid at different times: $t = 10^4 \, \rm yr$ ({\sl left panel}), $t = 10^5 \, \rm yr$ ({\sl middle panel}), and $t = 10^6 \, \rm yr$ ({\sl right panel}). The middle panel reproduces the data displayed in Fig.\,\ref{fig:paramgrid}, without the solutions with an ammonia binding energy of 1000\,K. The red lines represent our fiducial model, and the dark blue dashed lines represent the abundance profile deduced by \citet{Crapsi07}.}
    \label{fig:timedependence}
\end{figure*}

The results presented above display a clear discrepancy with the observations of \citet{Crapsi07}. From our models we always obtain an abundance profile that decreases strongly towards the center of the core. However, one further crucial aspect that we have not explored thus far is time-dependence. We plot in Fig.\,\ref{fig:timedependence} the results from our parameter grid at three different time steps.

The ammonia depletion timescale is very short in the central areas of the core because of the high density. Even at $t = 10^4 \, \rm yr$, which is an unrealistically short timescale for L1544 given that it already displays clear contraction motions, ammonia is heavily depleted in the innermost few thousand~au. A few thousand au away from the center, the solutions with chemical desorption included are already at least an order of magnitude above the observations in peak ammonia abundance, while the solutions without chemical desorption fall short of the observations by about an order of magnitude. In the outer core the difference of the model results as compared to the observed profile is greater still. If the gas is let to evolve to $t = 10^6 \, \rm yr$, none of the solutions show an acceptable agreement with the observations even when chemical desorption is excluded from the model.

We can deduce from the results presented in this Section that: 1) we cannot obtain with our physical and chemical models an ammonia abundance profile that does not present heavy ammonia depletion in the central areas of L1544; 2) the observations of \citet{Crapsi07} can be reproduced to a satisfactory degree, except inside the central few thousand au, if we exclude chemical desorption from the modelling, assume a sufficient amount of visual extinction in the parent cloud, and retain the canonical value for the CR ionization rate.

\section{Discussion}\label{s:discussion}

\subsection{Ammonia line emission simulations}
 
The comparison between the models and the observations presented above does not take into account any optical depth or excitation effects, which may affect for example the determination of the abundance from the observations. Therefore, the abundance profiles calculated here may not exactly correspond to the one derived by Crapsi et al. A more rigorous method of comparing models and observations is the reproduction of the emission lines with radiative transfer methods. To alleviate the ambiguity related to optical depth or excitation effects, we simulated the observed para-$\rm NH_3$ (1,1) inversion line and the ortho-$\rm NH_3$ ($1_0 - 0_0$) ground-state line with the non-local-thermal-equilibrium (LTE) radiative transfer code Cppsimu \citep{Juvela97}. For the ortho-line calculations we used collisional coefficients from \citet{Bouhafs17} which take the hyperfine splitting explicitly into account. Similar hyperfine-split collisional rates are not yet available for para-$\rm NH_3$, and so for the para-line calculations we used the data of \citet{Maret09} along with the assumption that the hyperfine components are split according to LTE.

\begin{figure}
	\includegraphics[width=1.0\columnwidth]{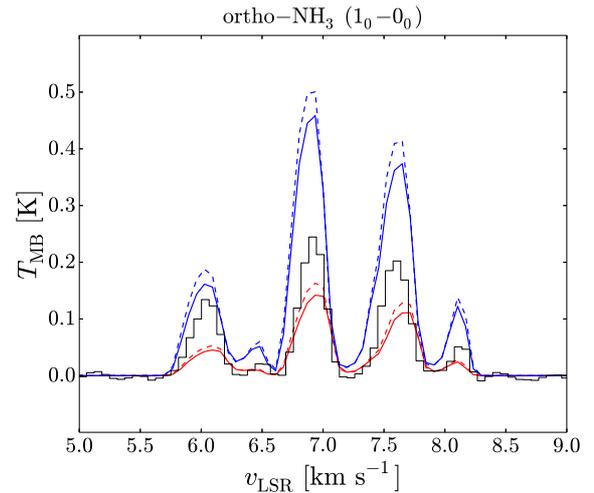}
    \caption{Observed ortho-$\rm NH_3\,(1-0)$ transition from \citet{Caselli17} (black line). The blue solid and dashed lines represent line simulations adopting $\rm NH_3$ ortho/para ratios of 0.7 and 1, respectively, using the physical structure of \citet{Crapsi07}. The red lines show the corresponding results using the \citet{Keto10a} physical structure. In both cases, the parametrized ortho-$\rm NH_3$ abundance profile is taken from \citet{Crapsi07}.}
    \label{fig:comparison}
\end{figure}

To test our radiative transfer setup, we attempted to reproduce the observations of \citet{Caselli17} using two different source profiles. First, we took the density, temperature, and $\rm NH_3$ abundance profiles given by \citet{Crapsi07} and used them as input to Cppsimu. The infall velocity profile was taken from K10, but multiplied by a factor of 1.75 \citep{Bizzocchi13}. Second, we took the physical structure from K10 and mapped the ammonia abundance profile from \citet{Crapsi07} onto this physical model, i.e., both models use the same parametrization for the ammonia abundance. The beam FWHM and spectral resolution were set to 40$\arcsec$ and 64\,m\,s$^{-1}$, respectively, corresponding to the {\sl Herschel} observations of \citet{Caselli17}. The results of this comparison are shown in Fig.\,\ref{fig:comparison}. We cannot reproduce the observed profile with either one of the two source structures. We point out that MOLLIE \citep{Keto90, Keto10b}, which was used for the line simulations presented in \citet{Caselli17}, is able to match the observation. We explore this issue further in Appendix~\ref{a:benchmark}, where we compare the results from the two codes in a couple of test cases. In what follows, we use Cppsimu.

\begin{figure}
	\includegraphics[width=1.0\columnwidth]{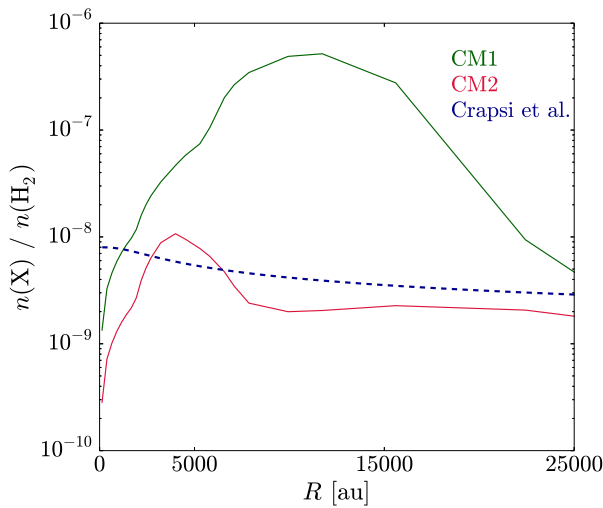}
    \caption{Best-fit abundance profile for ortho+para ammonia from model~CM2 (red), obtained at $t = 1.74 \times 10^5 \, \rm yr$. The corresponding abundance profile from model~CM1 at the same time step is shown in green. The dark blue dashed line represents the abundance profile deduced by \citet{Crapsi07}.}
    \label{fig:bestfit_abundances}
\end{figure}

To illustrate the emission lines associated with the abundance profiles presented in this paper, we used two different chemical schemes: 1) our fiducial model, and 2) the fiducial model with chemical desorption turned off and external $A_{\rm V}$ set to 5~mag. The latter model was chosen on the grounds that it provides a decent fit to the observed ammonia abundance profile (outside the core center) as discussed above. From here on we refer to these two models simply as CM1 and CM2. We considered 51 time steps logarithmically evenly spaced between $10^4$ and $10^6$\,yr and searched for the closest match to the observed abundance profile in model CM2 using a $\chi^2$ analysis. The best-fit abundance profile as determined by this analysis ($t = 1.74 \times 10^5 \, \rm yr$) is shown in Fig.\,\ref{fig:bestfit_abundances}.

\begin{figure*}
	\includegraphics[width=2.0\columnwidth]{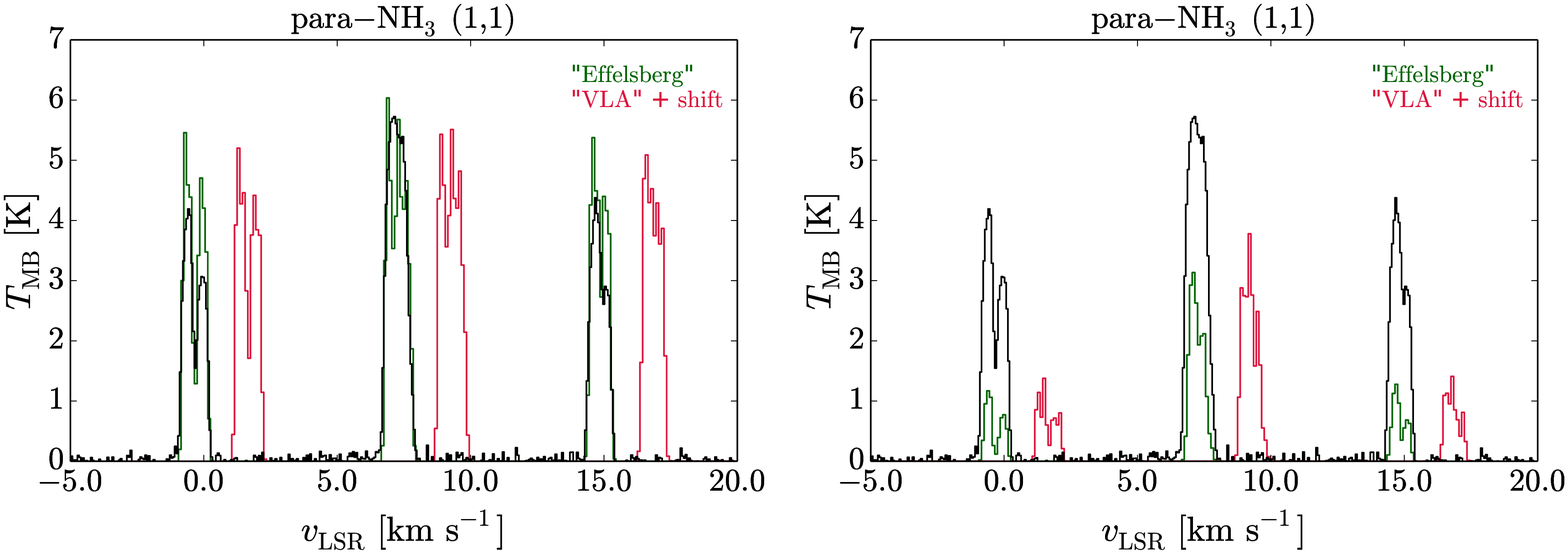}
    \caption{Radiative transfer simulations of the para-ammonia (1,1) line. The ammonia abundance profiles are taken from model~CM1 ({\sl left panel}; see text) or model~CM2 ({\sl right panel}). The simulated lines correspond to the single-dish Effelsberg 100m observations of \citet{Tafalla02} with a beam Full Width at Half Maximum (FWHM) of 37$\arcsec$ (green lines), and to the VLA observations of \citet{Crapsi07} which are here simulated with a circular beam with an FWHM of 4$\arcsec$ (red lines). The simulated VLA lines are offset by 2\,km\,s$^{-1}$ so that the differences in the spectra are clearer. The black lines reproduce the Effelsberg observations of \citet{Tafalla02}.}
    \label{fig:linesimulations_para}
\end{figure*}

Fig.\,\ref{fig:linesimulations_para} shows the results of the radiative transfer calculations for para-$\rm NH_3$. The Local Standard of Rest (LSR) velocity of L1544 is 7.2\,km\,s$^{-1}$ \citep{Tafalla98}. The spectral resolution of the simulation was set to $0.1\,\rm km\,s^{-1}$, and we limited the LSR range in the figure to encompass the three central components of the line as in Fig.\,2 in \citet{Crapsi07}. The critical density of the (1,1) transition is $\sim2.0\times10^3\,\rm cm^{-3}$ at 10\,K, which means that the line can be collisionally excited in a broad region inside $\sim$25000\,au (see Fig.\,\ref{fig:physicalmodels}). This implies that the (1,1) inversion transition does not probe the central parts of the core well. If we choose model CM1, the beam size does not play a large role because the ammonia abundance is high and hence the column towards the core center is large even when smoothed to 37$\arcsec$. If we instead choose model CM2 where the ammonia abundance profile more or less follows the one derived by \citet{Crapsi07} -- except in the center of the core -- the effect of the beam size is accentuated.

\begin{figure}
	\includegraphics[width=1.0\columnwidth]{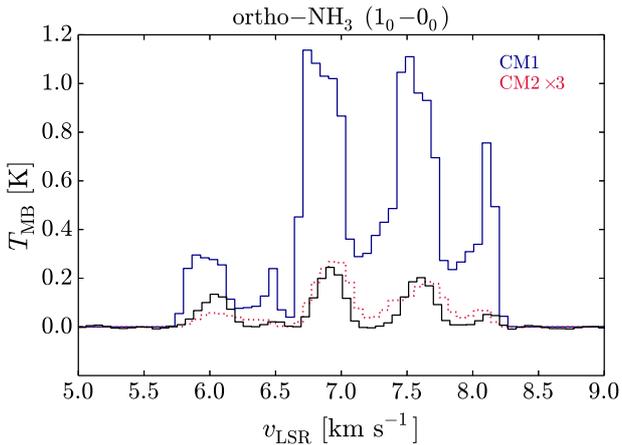}
    \caption{Radiative transfer simulations of the ortho-ammonia ($1_0-0_0$) line. The simulated lines correspond to the {\sl Herschel} observations of \citet{Caselli17}, shown in black. The ammonia abundance profiles are taken from model~CM1 (blue; see text) or model~CM2 (red). The spectrum from model~CM2 has been multiplied by 3.}
    \label{fig:linesimulations_ortho}
\end{figure}

Neither one of the chemical models provides a good fit to the observations of \citet{Tafalla02} and \citet{Crapsi07}. Model~CM1 reproduces the main peak intensity of the Effelsberg observations but the satellite lines are too strong, which may also be due to the approximation of LTE for the distribution of the hyperfine components. The optical thickness in the simulations is clearly higher than observed. The VLA observations are not reproduced with this model either, as the intensities of the hyperfine components are overestimated by about 2\,K compared to the observations (unfortunately the spectra of \citet{Crapsi07} are not available to facilitate easy comparison). Model~CM2 spectacularly fails to reproduce either observation owing to the missing ammonia in the innermost 5000\,au. This shows that even though the (1,1) transition does not probe the core center specifically, emission from the central areas is still crucial when the ammonia abundance in the outer core is low.

Fig.\,\ref{fig:linesimulations_ortho} shows the simulated lines for ortho-$\rm NH_3$. The critical density of the ($1_0-0_0$) line at 10\,K is $\sim 3.7 \times 10^7 \, \rm cm^{-3}$, and so the line is collisionally excited only in the dense inner part of the core. This feature is especially evident in model~CM2 where the ammonia abundance is low in the central area and the resulting emission is very faint. Model~CM1 however produces a line that is much too bright and optically thick compared to the observations, and so once again the two chemical models fail to reproduce the observations.

Model~CM2 is one example case in a family of solutions that is close to the abundance profile derived by \citet{Crapsi07}, except in the inner core. While we did not calculate $\chi^2$ values for all of our models, it is evident from Figs.\,\ref{fig:paramgrid}~and~\ref{fig:singlemodels} and from the analysis presented above that we do not obtain a solution that could reproduce the observed ammonia line profiles, even taking into account possible calibration errors in the observations. {\it The presence of ammonia in the gas phase in the high-density central regions of the core is required}.

\subsection{Distributions of chemical species related to ammonia formation}

\begin{figure}
\centering
	\includegraphics[width=0.9\columnwidth]{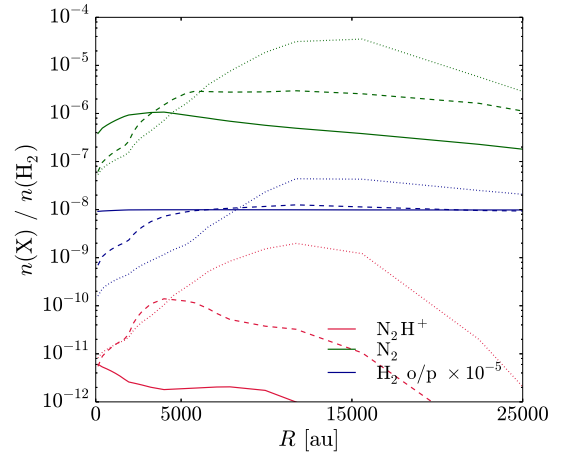}
    \caption{Distributions of $\rm N_2H^+$ ({\sl red}), $\rm N_2$ ({\sl green}), and $\rm H_2$ o/p ratio ({\sl blue}; multiplied by $10^{-5}$ for better readability of the plot) in our fiducial model at $t = 10^4 \, \rm yr$ ({\sl solid lines}), $t = 10^5 \, \rm yr$ ({\sl dashed lines}), and $t = 10^6 \, \rm yr$ ({\sl dotted lines}).}
    \label{fig:otherspecies}
\end{figure}

As already alluded to in Sect.\,\ref{ss:singlemodels}, the gas-phase formation efficiency of ammonia depends on the abundance of $\rm N_2$: the reaction $\rm N_2 + He^+ \longrightarrow N^+ + N + He$ produces $\rm N^+$ which can then be converted to $\rm NH^+$ through $\rm N^+ + H_2 \longrightarrow NH^+ + H$. The latter reaction is heavily dependent on the $\rm H_2$ ortho/para ratio because it is strongly endothermic when the reaction partner is para-$\rm H_2$, but close to thermoneutral with ortho-$\rm H_2$ \citep{Dislaire12}. Therefore, ammonia formation should be the most efficient when there is a large amount of $\rm N_2$ in the gas and when the $\rm H_2$ ortho/para ratio is high. $\rm N_2$ itself cannot be observed in pre-stellar cores, but $\rm N_2H^+$ can be used as a proxy of its abundance distribution. We plot in Fig.\,\ref{fig:otherspecies} the distributions of $\rm N_2H^+$ and $\rm N_2$, and the $\rm H_2$ ortho/para ratio, in our fiducial model at three different time steps. $\rm N_2$ freezes out onto the grain surfaces in the central core in a relatively short timescale, and this behavior is clearly reflected in the abundance of $\rm N_2H^+$. We note that the shape of the $\rm N_2H^+$ abundance profile does not strictly follow that of $\rm N_2$, because $\rm N_2H^+$ formation requires $\rm H_3^+$ which in turn reacts preferentially with CO while it is still available in the gas phase. The initial $\rm H_2$ ortho/para ratio is already low ($10^{-3}$; Table~\ref{tab:initialabundances}) which means that the production of $\rm NH^+$ and hence of ammonia is hindered. The ratio can go as low as $\sim 10^{-5}$ in the central core. It is possible that we are underestimating the initial $\rm H_2 \, \rm o/p$ ratio, but the single model S3 shows that the initial value has little bearing on the ammonia abundance at high density and low temperature.

The $\rm H_2$ o/p ratio that our model predicts is in line with previous observations and models \citep[for targets other than L1544; e.g.,][]{Brunken14,Furuya16,Bovino17}, and hence there is little reason to believe that an unexpectedly high o/p ratio would be present in the center of L1544, boosting the formation of ammonia beyond the levels obtained with our current model. We also note that $\rm N_2H^+$ and, to a lesser extent, $\rm N_2D^+$ depletion has been recently observed in L1544 (Redaelli et al., subm.). It is all the more puzzling why some species related to $\rm N_2$ show signs of freeze-out ($\rm N_2H^+$), while others do not (ammonia).

\subsection{Outlook on future modelling efforts}

Our chemical model does not take into account some effects that may influence the ammonia abundance. One of these is the treatment of multilayer ice chemistry coupled with dynamic chemical desorption efficiencies depending on the composition of the surface, such as in the recent study of \citet{Vasyunin17}. Their description of chemical desorption is based on the work of \citet{Minissale16a}, who unfortunately did not estimate desorption efficiencies for the reactions involved in the ammonia formation through hydrogenation (starting from N + H). While we cannot make a reliable test of the effect of chemical desorption on the relevant hydrogenation reactions owing to the lack of quantitative experimental and theoretical data, we did test the effect of the $\rm N + N \longrightarrow N_2$ reaction in our fiducial model by setting the desorption efficiency to the theoretical value of 89\% (for bare grains) given by \citet{Minissale16a}. We find that the N + N reaction is too slow for the enhanced $\rm N_2$ chemical desorption to be of consequence, and indeed the influence of this change on the gas-phase ammonia abundance is negligible. A more complete test can be carried out once experimentally or theoretically derived chemical desorption efficiencies for the appropriate hydrogenation reactions become available.

Another issue is the question of dynamic binding energies on the grain surface. At early times the grains will be covered mostly with water ice, but later as CO starts to freeze out, the grains will be covered with CO and other species. This will change the binding energies of the various species on the surface, and since ammonia is a late-type molecule, it is conceivable that the inclusion of this effect would lead to a decreased depletion factor for ammonia since the binding energy of ammonia on an apolar molecule such as CO will be lower than on water. It is however difficult to formulate this issue in a chemical model in a physically meaningful way, and experimental data is also lacking. Further studies, both theoretical and experimental, are certainly called for.

We demonstrated that considering a self-consistent hydrodynamical treatment of the core collapse coupled with chemistry and radiative transfer (an update of the model presented in \citealt{Sipila18}) does not solve the problem of strong ammonia depletion in the central core. Nevertheless it is evident that the dynamics needs to be taken into account in order to track the time-evolution of the abundances of the various species properly.

L1544 is a well-studied object that presents clear contraction motions and no signs of a central source, which implies that it is a pre-stellar core in its final stages of evolution towards becoming a protostellar system. We therefore expect the central area of the object to be cold, and since it is also well-shielded from external radiation which could contribute to the chemistry via photodesorption for example, we come to the conclusion that the inability of the present chemical model to reproduce the observed ammonia abundance is due to the still limited understanding of the chemical and physical processes at play. Time-dependent models present the most promising avenue of further study into this issue. We note that especially large uncertainties pertain to the binding energies of the various species on different types of ice, and that more experimental and theoretical work on this problem is urgently needed so that chemical models of interstellar chemistry can provide reliable results.

\section{Conclusions}\label{s:conclusions}

We performed a parameter-space exploration of the abundance of (ortho+para) ammonia predicted by gas-grain chemical models with different assumptions of parameter values and included/excluded processes. Our goal was to determine if the abundance profile of ammonia observed toward the pre-stellar core L1544, which shows an inward-increasing trend, can be reproduced by considering variations of standard chemical model parameters. For the sake of simplicity and to keep the computational time at a reasonable level, we used in most of our models a static physical structure for L1544 from K10. We varied six different modelling parameters (such as the ammonia binding energy and the sticking coefficient of atomic N) that a priori should influence the ammonia abundance in a significant way, resulting in a total of 108 models. We also considered eleven other models in which we tested the influence of other important parameters such as the initial $\rm H_2$ ortho/para ratio, and the effect of dynamics.

Observations of the ammonia abundance toward L1544 by \citet{Crapsi07} indicate that the abundance increases monotonically toward the center of the core. We found that irrespective of the various parameters we cannot obtain such a profile with our models. The various parameter combinations yield results with varying degrees of ammonia depletion in the central area of the core depending also on the time, but the depletion always occurs -- even in a purely gas-phase model, where the effect is due to processes other than adsorption. Interestingly, the models where chemical desorption (which is here modeled assuming a uniform 1\% efficiency) is taken into account are seemingly ruled out by our results, as these models lead to solutions where the ammonia abundance in the outer core is orders of magnitude above the observed one. This would constrain the chemical desorption efficiency to below 1\%, at least for ammonia on water ice. We also confirmed with radiative transfer simulations that the emission lines arising from our modeled abundance profiles cannot match the observed ones.

Our results point toward a dynamic nature of the chemistry and of the underlying physical processes. On the one hand, static physical models naturally cannot account for the abundance variations caused by time-evolution of the core density profile, and this will be reflected on observable emission lines \citep{Sipila18}. On the other hand, considering the effect of the dynamically-varying chemical composition of the ice surface can have a great impact on the efficiency of chemical desorption, as \citet{Vasyunin17} have recently shown. In addition to chemical desorption, dynamically-varying chemical abundances affect a multitude of other processes, such as line cooling and self-shielding of molecules such as $\rm H_2$, CO, and $\rm N_2$. It is increasingly evident that simplified pseudo-time-dependent models of interstellar chemistry provide only a limited explanation of the chemical complexity that is observed in the ISM, and that future modelling efforts should concentrate on time-dependent effects. Also, laboratory measurements of the binding energy of ammonia onto ices with variable compositions are required.

\section*{Acknowledgements}

We thank the anonymous referee for valuable comments. M.J. acknowledges the support of the Academy of Finland Grant No. 285769. P.C. acknowledges the support of the ERC Advanced grant PALs 320620.

%%%%%%%%%%%%%%%%%%%%%%%%%%%%%%%%%%%%%%%%%%%%%%%%%%


\bibliographystyle{mnras}
\bibliography{refs} % if your bibtex file is called example.bib

%%%%%%%%%%%%%%%%%%%%%%%%%%%%%%%%%%%%%%%%%%%%%%%%%%

%%%%%%%%%%%%%%%%% APPENDICES %%%%%%%%%%%%%%%%%%%%%

\appendix

\section{Radiative transfer code benchmark}\label{a:benchmark}

\begin{figure*}
	\includegraphics[width=1.7\columnwidth]{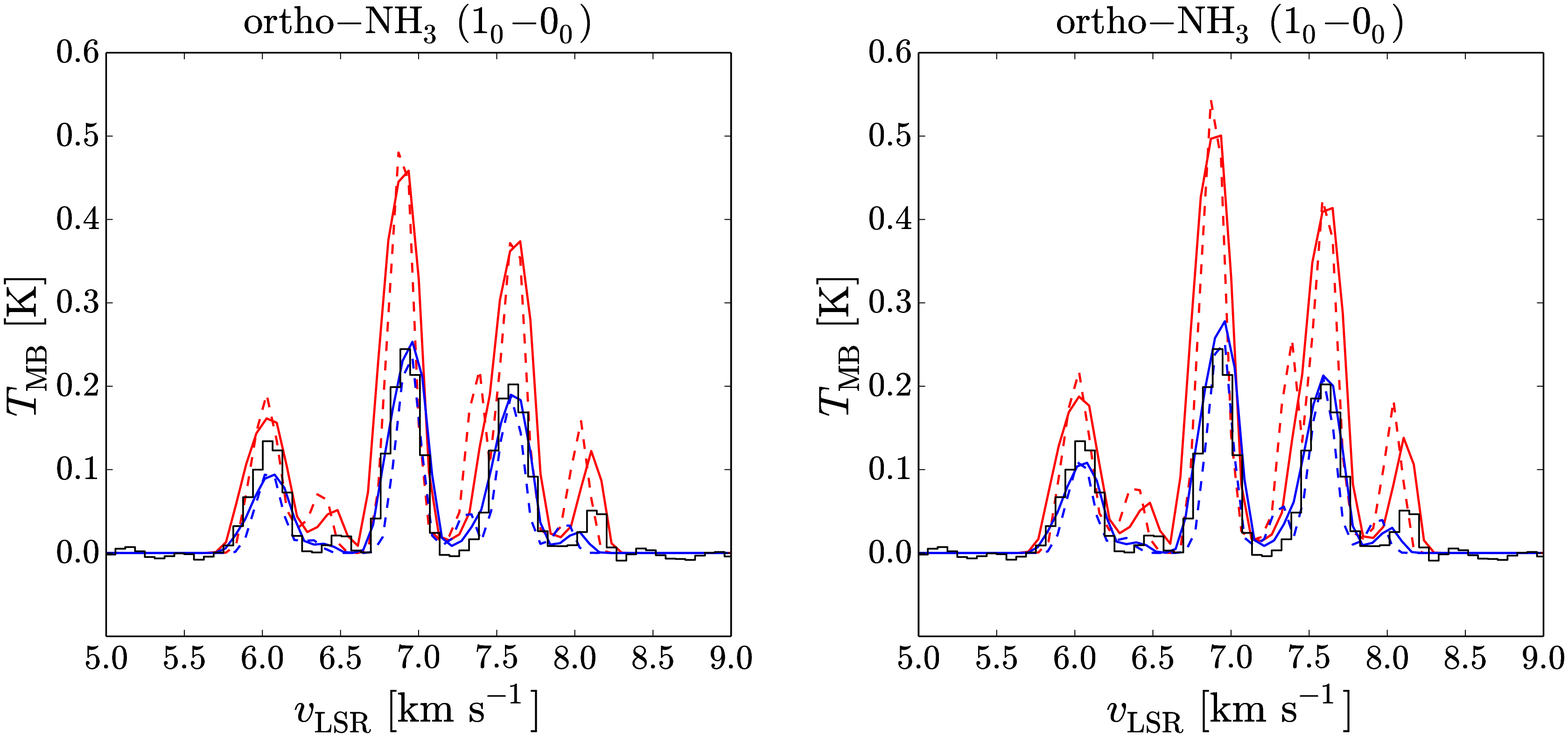}
    \caption{Line emission simulations for the ortho-$\rm NH_3\,(1-0)$ transition using Cppsimu (red lines) and MOLLIE (blue lines). The physical source model is taken from \citet{Crapsi07}. Solid lines correspond to solutions where the \citet{Keto10a} velocity profile has been scaled by a factor of 1.75, while the dashed lines correspond to no scaling. The left and right panels correspond to an $\rm NH_3$ ortho/para ratio of 0.7 or 1.0, respectively.}
    \label{fig:comparison_crapsi}
\end{figure*}

\begin{figure*}
	\includegraphics[width=1.7\columnwidth]{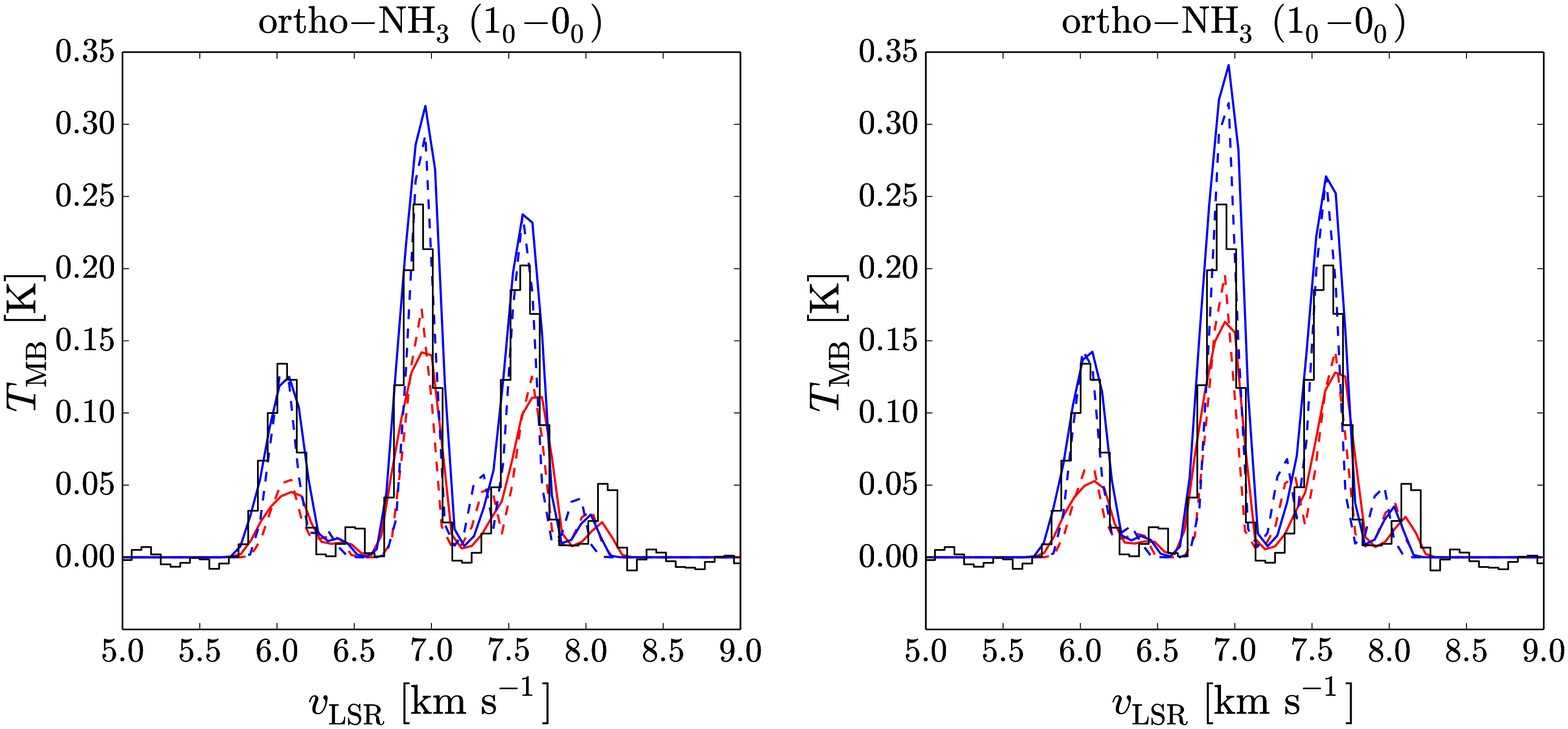}
    \caption{As Fig.\,\ref{fig:comparison_crapsi}, but the physical source model is taken from \citet{Keto10a}.}
    \label{fig:comparison_keto}
\end{figure*}

Here we compare simulated emission lines calculated with Cppsimu or MOLLIE. We focus first on the ortho-$\rm NH_3$ ground-state rotational transition. Figure~\ref{fig:comparison_crapsi} shows the results for the \citet{Crapsi07} physical structure (i.e., medium density and gas temperature), and Fig.~\ref{fig:comparison_keto} shows the corresponding calculation for the K10 physical structure. The \citet{Crapsi07} density structure, which is given as a parametrized formula, was set so that it corresponds to the same outer radius as the K10 model. In both cases the ortho-$\rm NH_3$ abundance profile was taken from \citet{Crapsi07}. The infall velocity profile was adopted from K10; we also studied the effect of scaling up the velocity profile by a factor of 1.75 \citep[see][]{Bizzocchi13}.

MOLLIE fits the observed line profile fairly well when the \citet{Crapsi07} physical structure is used, whereas Cppsimu overestimates the brightness of the strongest component by almost a factor of two. Curiously, when we switch to the K10 physical structure, MOLLIE overestimates the emission while Cppsimu underestimates it. An extra feature at $v_{\rm LSR} = 7.4 \, \rm km \, s^{-1}$ that is not seen in the observations is predicted by both codes when the velocity field of K10 is not scaled up. Despite the evident discrepancy, the results from the two codes are within factor of two.

\begin{figure*}
	\includegraphics[width=2.0\columnwidth]{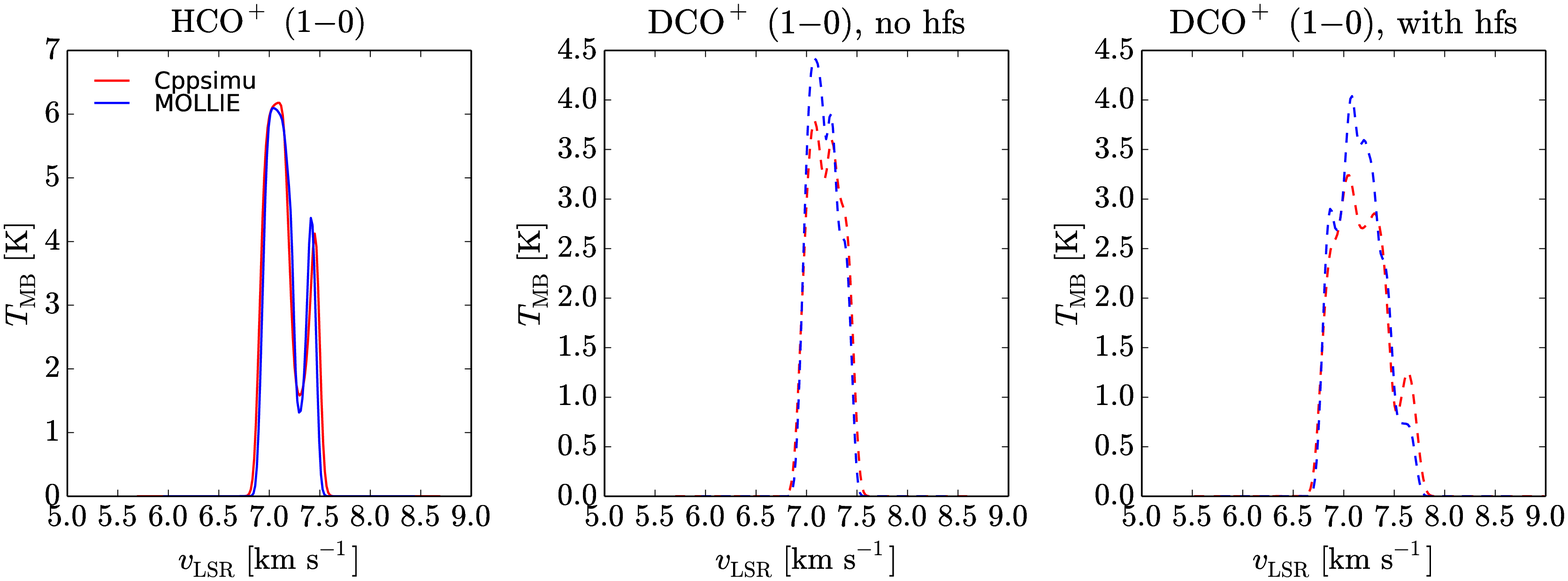}
    \caption{Comparison of the $\rm HCO^+$ (1-0) line and the $\rm DCO^+$ (1-0) line with and without hyperfine structure (hfs), calculated with Cppsimu (red) or MOLLIE (blue).}
    \label{fig:comparison_hco+}
\end{figure*}

We also compared the line emission profiles of the (1-0) rotational transitions of $\rm HCO^+$ and $\rm DCO^+$, excluding or including hyperfine structure for $\rm DCO^+$ (Fig.\,\ref{fig:comparison_hco+}). The K10 physical structure (with 1.75 velocity scaling) was used for these tests, and we took the appropriate spectral and collisional data from LAMDA \citep{Schoier05}. The $\rm HCO^+$ line profiles match well between the two codes, but for $\rm DCO^+$, MOLLIE produces brighter lines, as already seen with ortho-$\rm NH_3$ when using the K10 physical structure. The difference in line brightness does not appear to depend on the inclusion or exclusion of hyperfine structure, indicating that the differences in the ammonia spectra provided by Cppsimu and MOLLIE are not due to an incompatible treatment of hyperfines.

We have not found an obvious cause for the discrepancies in the radiative transfer simulations, although it is difficult to model well highly optically thick lines with non-LTE radiative transfer codes \citep[see also][]{Quenard16}. It is nevertheless very unlikely that our general conclusions on the modeled vs. observed lines, presented in the main body of the paper, would be affected by a different choice of the radiative transfer code used to produce the simulated lines (i.e., adopting a code other than Cppsimu or MOLLIE). Based on our tests we cannot conclude that MOLLIE yields the ``correct'' result, given that the positions and relative strengths of some spectral features are better reproduced by Cppsimu. Also, the MOLLIE results are counterintuitive in that one would -- naively -- expect weaker emission in the high-critical-density ortho-ammonia line when using the K10 physical model which is highly centrally concentrated (owing to self-absorption), while MOLLIE predicts the opposite.

%%%%%%%%%%%%%%%%%%%%%%%%%%%%%%%%%%%%%%%%%%%%%%%%%%

% Don't change these lines
\bsp	% typesetting comment
\label{lastpage}
\end{document}